\documentclass[prl,twocolumn,showpacs,amsmath,amssymb,sortaddress,superscriptaddress]{revtex4-1}
\usepackage{graphicx}
\usepackage{dcolumn}
\usepackage{colortbl}   
\newcommand{\br}{{\bf r}}                                                     
                                                     
\newcommand{\bq}{{\bf q}}  
\newcommand{\bs}{{\bf s}}

\newcommand{\bK}{{\bf K}}     
\newcommand{\bP}{{\bf P}}       
       
\newcommand{\ie}{\begin{equation}}                                             
\newcommand{\fe}{\end{equation}}                                               
\newcommand{\iea}{\begin{eqnarray}}                                            
\newcommand{\fea}{\end{eqnarray}}

\newcommand{\LSCO}{La$_{2-x}$Sr$_{x}$CuO$_{4}$}

\begin{document}
\title{Electronic polymers and soft-matter-like broken symmetries in
  underdoped cuprates}

\author{M. Capati}
\affiliation{Dipartimento di Fisica, Universit\`a di Roma Sapienza, Piazzale Aldo
Moro 5, I-00185 Roma, Italy}
\affiliation{ISC-CNR, Via dei Taurini 19, I-00185 Roma, Italy}

\author{S. Caprara}
\author{C. Di Castro}
\author{M. Grilli}
\affiliation{Dipartimento di Fisica, Universit\`a di Roma Sapienza, Piazzale Aldo
Moro 5, I-00185 Roma, Italy}
\affiliation{ISC-CNR, Via dei Taurini 19, I-00185 Roma, Italy}
\affiliation{CNISM Unit\`a di Roma Sapienza, Piazzale Aldo
Moro 5, I-00185 Roma, Italy}

\author{G. Seibold}
\affiliation{Institut f\"ur Physik, BTU Cottbus - Senftenberg, PBox 101344, 03013 Cottbus, Germany}

\author{J. Lorenzana}
\affiliation{Dipartimento di Fisica, Universit\`a di Roma Sapienza, Piazzale Aldo
Moro 5, I-00185 Roma, Italy}
\affiliation{ISC-CNR, Via dei Taurini 19, I-00185 Roma, Italy}

\date{\today}
\begin{abstract}
{\bf  
Empirical evidence in heavy fermion, pnictide, and 
other systems suggests that unconventional superconductivity 
appears associated to some form of real-space electronic
order. For the cuprates, despite several proposals, the
emergence of order in the phase diagram between the commensurate antiferromagnetic state and the
superconducting state is not well understood. Here we show that in
this regime doped holes assemble in ``electronic 
polymers''. Within a Monte Carlo study we find, that in clean systems by lowering the temperature 
the polymer melt condenses first in a smectic state and then in a Wigner crystal both with the addition 
of inversion symmetry breaking. Disorder blurs the positional order leaving a robust inversion symmetry 
breaking and a nematic order, accompanied by vector chiral spin order and
with the persistence of a thermodynamic transition. Such electronic
phases, whose  properties are reminiscent of soft matter physics, produce charge and spin responses in good accord with 
experiments.
}
\end{abstract}

\pacs{74.25.Ha, 71.28.+d, 75.25.-j}

\maketitle

\vspace{1 truecm}
\noindent
{\bf\large Introduction}

The anomalous behavior of several physical 
quantities above the superconducting transition temperature 
suggests that high-temperature superconductivity emerges close to a quantum
critical point between a broken-symmetry state and the disordered
phase, in analogy with heavy fermion materials, pnictides and organics~\cite{Taillefer_ARCMP2010}.
However, identifying the broken-symmetry phases has proven much more difficult than in other 
materials, despite several ``gold rushes'' in the underdoped region of the phase diagram, 
triggered by the observation of stripe 
order~\cite{Tranquada_Nature1995,Tranquada_PRL1997}, 
nematic order~\cite{Hinkov_Science2008,Haug_NJOP2010,Daou_Nature2010}, 
time-reversal symmetry breaking~\cite{Fauque_PRL2006}, and incommensurate charge-density-wave
order~\cite{Ghiringhelli_Science2012}.   
  
While some form of charge order (CO) is well 
established~\cite{vojta2009_review}, an important question is how this order 
is formed starting from the two extremes of the phase diagram. When coming from the high-doping region, 
CO seemingly arises as a second-order instability of the uniform {\em strongly correlated} metallic 
state, producing incommensurate charge density waves driven by 
magnetic~\cite{Sachdev_PRL2013,Wang_PRB2014}, 
phononic~\cite{Castellani_PRL1995,Andergassen_PRL2001}, 
or mixed~\cite{Castellani_JPCS1998} microscopic mechanisms. 
On the other hand, the occurrence of CO 
from the Mott insulating low-doping side is more directly tied to the
tendency of Mott antiferromagnets to expel and segregate
charges~\cite{Emery_PRL1990,Marder_PRB1990,Cancrini_EPL1991,Castellani_PRB1991}.This latter region is the  playground of our present work. 

Recently, it was pointed out~\cite{Seibold_PRB2013,Seibold_SR2014} 
that, at very low doping, charge segregation may 
acquire features, related to the occurrence of topological excitations in doped antiferromagnets. 
The starting point is the observation that holes in an antiferromagnet induce a vortex (V) or 
antivortex (A) texture in the surrounding spin ordering. While isolated vortices are energetically 
expensive, a VA pair is 
stable~\cite{Seibold_PRB1998,Verges_PRB1991,Berciu_PRB2004,Koizumi_JPSJ2008}, 
because its annihilation is hindered by the strongly 
correlated character of the doped holes and the disturbance of the antiferromagnetic 
background rapidly dies out at large distances. As in early
proposals~\cite{Aharony_PRL1988,Timm_PRL2000} inspired by the work of 
Villain~\cite{Villain_JPC1977,Villain_JPC1977b},
these VA ``dimers'' explain the extremely rapid  destruction of
long-range antiferromagnetic order with doping. 
\begin{figure}[tb]
\includegraphics[width=8.5cm]{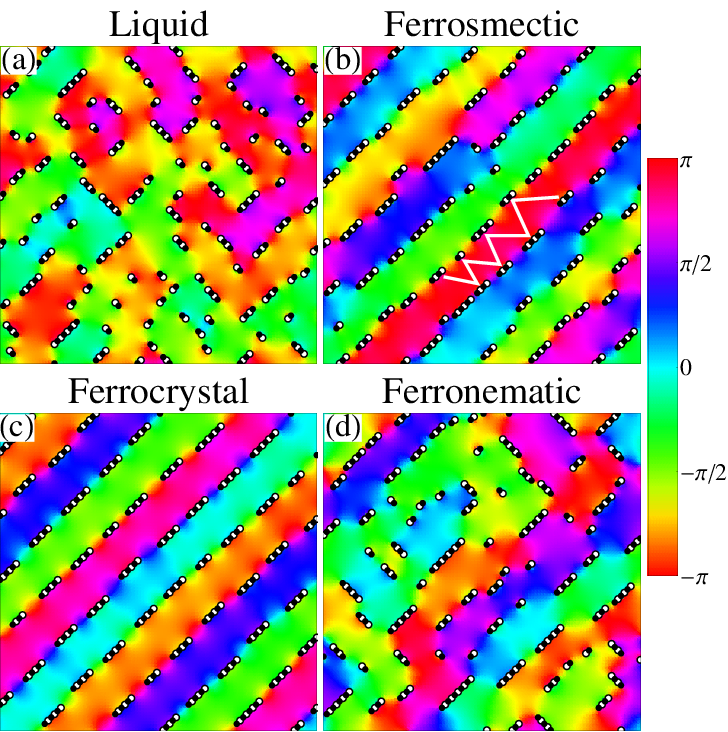}
\caption{{\bf Charge and spin configurations in the
 different phases.}  White and black circles represent the positive 
and negative topological charges, respectively. The different colours
denote the angle of the staggered magnetization. The images are 
Monte Carlo snapshots in the absence of quenched disorder (a-c) in
the thermally disordered phase with $T=50$\,K (a), in the ferrosmectic
phase at $T=38$\,K (b), in the ferrocrystal phase at $T=8$\,K (c)
 and in the ferronematic phase at $T=40$\,K (d)
 which appears in the presence of quenched disorder ($Q_{\mathrm{ion}}/Q_{\mathrm{rep}}=0.125$) . The white lines in (b) highlight the ``triangular'' arrangement of the segments.}
\label{fig:snap}
\end{figure}

In the present scenario the dimers or ``nematogens'' 
self-organize and give rise to ``electronic soft matter'' 
effects~\cite{Kivelson_Nature1998}. Specifically, the dimers 
may undergo a ``polymerization process'', triggering charge segregation into segments, tightly bound to V and A spin textures. These segments not only align forming a nematic state, but can also break 
inversion symmetry~\cite{Villain_JPC1977,Villain_JPC1977b}, 
due to their intrinsic topologic dipolar character (associated with the V and A at 
the endpoints of the ``polymer''). This state, which was named 
{\em ferronematic}~\cite{Seibold_PRB2013,Seibold_SR2014}, is 
accompanied by a spin spiral state sustaining a net spin current. 
At large scales, this feature is reminiscent of other 
proposals~\cite{Shraiman_PRL1989,Gooding_PRB1991,Hasselmann_PRB2004,Sushkov_PRL2005}, which are however based
on impurity states instead of the polymer states which are central to
our results. A ferronematic phase was proposed also to occur in ultra-cold
dipolar Fermi gases of atoms~\cite{Fregoso_PRL2009}.

We pose here the following fundamental questions: which other phases
can be sustained by the electronic polymers, how are they affected by
quenched disorder, what is the fate of the thermodynamic phase
transitions expected in ideally clean systems, and how their characteristic temperature scales
emerge from the (usually much higher) electronic scales of the
system. In order to study the problem at the large length scales probed
by experiments we carry out a multiscaling approach starting from a microscopic
model and derive a mesoscale effective model treated with Monte Carlo. We
obtain a rich phase diagram for the electronic polymers as a
function of temperature and disorder  
which allows to rationalize the charge and spin responses observed
experimentally.

\vspace{1 truecm}
\noindent
{\bf\large Results}

{\bf Numerical Simulations.}
We start from the very low doped regime of few  holes in the spin
antiferromagnetic background of a CuO$_2 $ plane modeled by a one-band
Hubbard model. We study the dimers at mean-field level in the
Gutzwiller approximation (GA) (see Methods, Supplementary Note 1 and
Supplementary Fig. 1). With realistic parameters for \LSCO, the most
favorable configuration for two holes is along the diagonal of a plaquette with a planar dipolar distortion of the antiferromagnetic 
background. The latter can be visualized as due to a V and an A
centered close to  (but not exactly at) the vertices of the
plaquette and forming a ``topological dipole'' (TD). 
There is another two-hole mean-field solution which is non-planar and consist
of a skyrmion texture~\cite{Seibold_PRB1998} which, for the present parameters, 
is $\sim$ 100~K higher in energy than the TD and
therefore will be neglected at low temperatures in the following.

Studying metastable planar configurations, 
in which two or more of these TDs are arranged with different positions 
and orientations (Supplementary Figs. 2 and 3)
 we find, as expected, that at large distances holes
interact through a logarithmic interaction~\cite{Minnhagen_RMP1987} between
their topological charges, while at short distances their interaction
is modified by quantum effects related to the  overlap of the hole
wave functions. The logarithmic interaction stems from the fact that
for planar textures the long-range behavior can be captured by an XY
model~\cite{Aharony_PRL1988,Timm_PRL2000}.  Notice that we are not claiming that the symmetry
of the model is reduced from Heisenberg to XY. Indeed each texture has
a zero mode related to the change of the plane which contains the
spins, as it should for an O(3) symmetric model. However, contrary to
what would happen for a single vortex in the pure Heisenberg model,
the textures have no unstable modes~\cite{Seibold_SR2014} which would break the
planar character of the texture, i.e. they are locally stable and
their energy is correctly captured by a planar magnetic model.

In order to enable simulations in large systems we do not consider explicitly 
the spin degrees of freedom but integrate them out to generate
effective interactions among topological charges. While this is an
enormous computational advantage, it limits our simulations to low
doping ($n_h\lesssim 0.1$ holes per unit cell)  where spin currents
are small on average and the superposition principle is valid,
allowing for a mapping of topological charges onto a two-dimensional
(2D) Coulomb gas~\cite{Minnhagen_RMP1987}. The effective interaction among
topological defects, needed for the Coulomb gas model, is obtained by
fitting the energy of several metastable zero-temperature GA 
solutions obtained  in the  Hubbard model.

As a consequence of the interaction mediated by the antiferromagnetic
background, when a large  even number of holes is added to the system, these tend to bind into a single polymeric chain of 
alternating topological charges, ending with a V and an A. Adding the real three-dimensional long-range 
Coulomb repulsion among holes, whose strength is measured by a parameter $Q_{\mathrm{rep}}$ (see Methods), these 
long polymers break into smaller polymers, as shown in
Fig.~\ref{fig:snap} and Supplementary Fig. 5.

We work at temperatures $T$ smaller than the binding energy of individual VA pairs ($\approx 100$\,K), so that 
the number of unbound topological charges is negligible. Therefore, our basic constituents in the Monte Carlo computations are the TDs. These are modeled by a bound V and A, each moving on the sites of a square 
lattice, with the topological charge adjusted so that the dipole moment matches the 
Gutzwiller computations (see Methods). Since there are no topological constraints on
the charge $\pm k$ of the V and the A, they turn out to be fractional,
$k\approx 0.8$ (see Supplementary Note 1).

A crucial problem in cuprates is to determine how disorder affects the ordered phases of 
the ideal ``clean'' system~\cite{Nie_PNAS2014}. In order to address this issue, the holes
attached to the topological charges are subject to a ionic disorder potential with strength  
$Q_{\mathrm{ion}}$, generated by the counterions out of the CuO$_2$
plane (see Methods and Supplementary Note 2 and Supplementary Fig.~4). The magnitude of 
$Q_{\mathrm{ion}}$ is difficult to estimate because it depends on screening processes not 
comprised in the model. Therefore, we treat $Q_{\mathrm{ion}}/Q_{\mathrm{rep}}$ as a phenomenological 
dimensionless parameter which characterizes the amount of disorder.

\begin{figure}[tb]
\includegraphics[width=8.5cm]{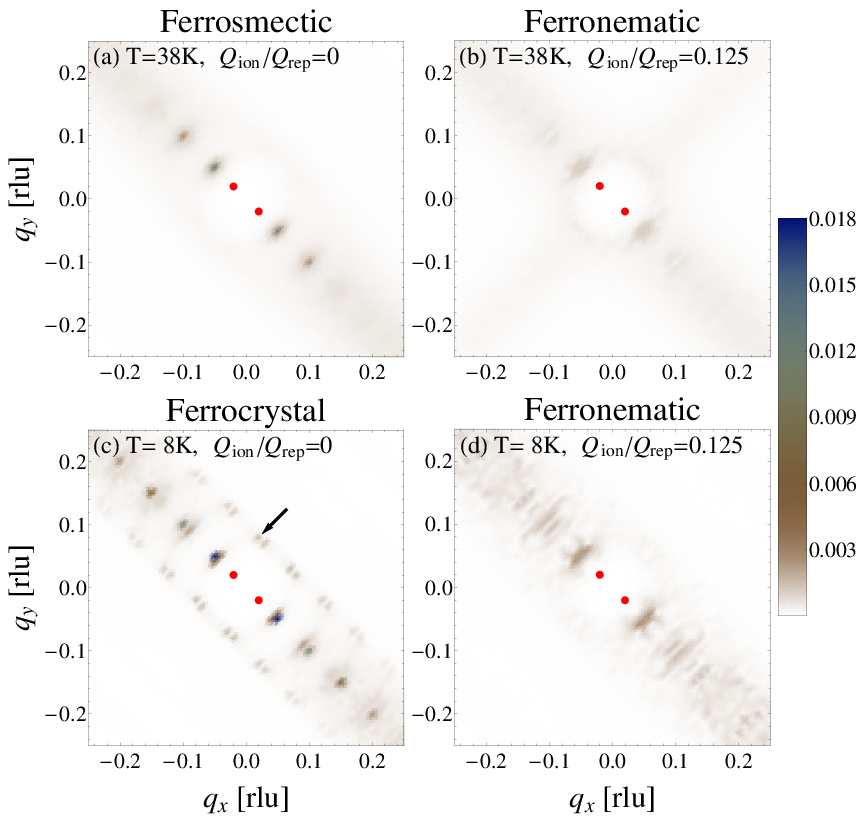}
\includegraphics[width=8.5cm]{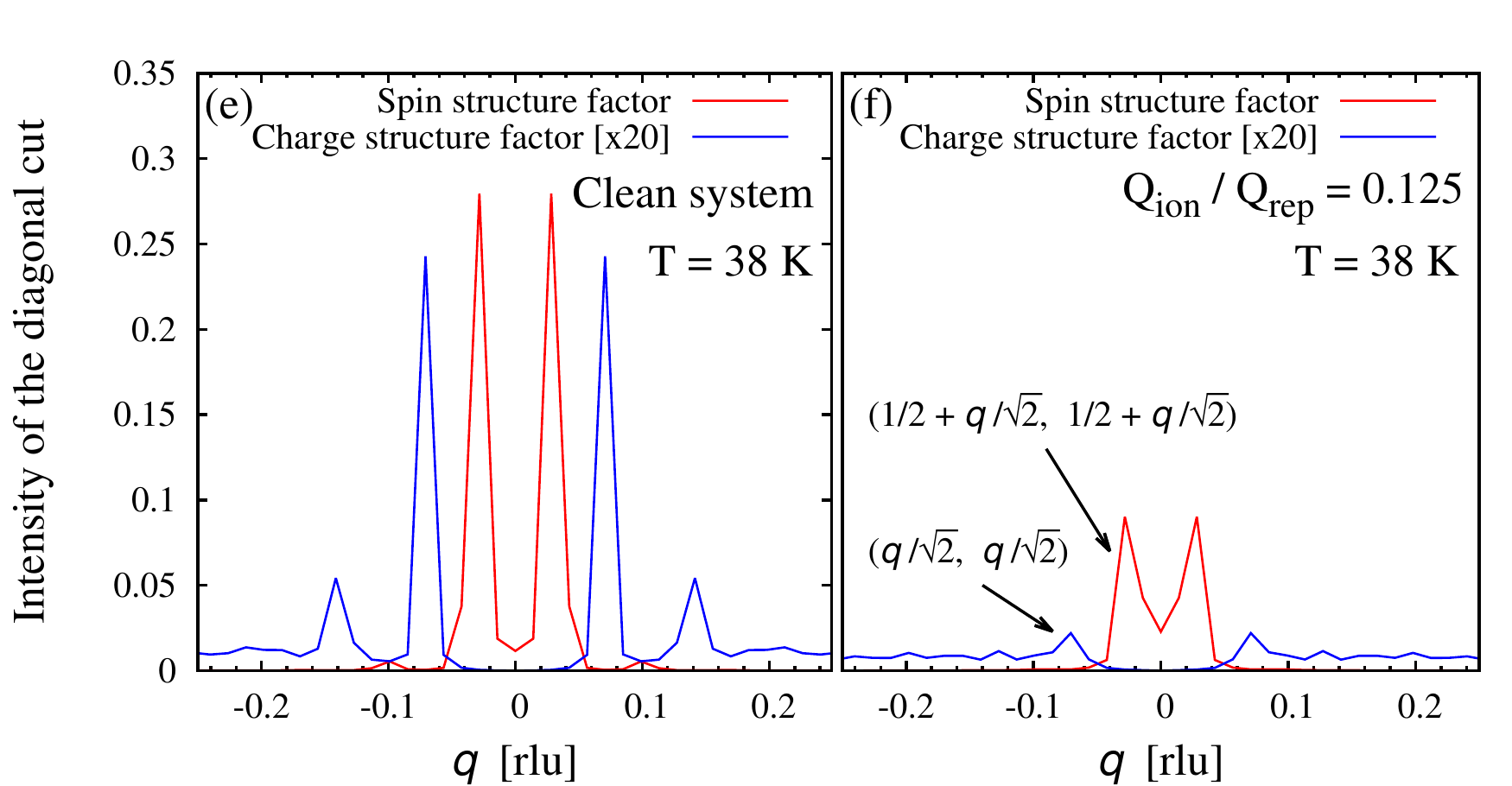} 
\caption{{\bf Charge and spin structure factor.}
Density plots in the 2D reciprocal space for: the charge structure factor for the clean system 
at (a) 38\,K and (c) 8\,K; the charge structure factor for the system with  $Q_{\mathrm{ion}} / Q_{\mathrm{rep}} = 0.125$ 
at (b) 38\,K and (d) 8\,K. The red solid circles represent the position of the spin peaks in the reciprocal 
space [shifted by $q_{AF}$]. The arrow in panel (c) shows a 
ferrocrystal peak. 
The lower panels show the diagonal cut of the charge and staggered spin 
structure factor in the 2D reciprocal space at 38\,K for (e) the clean system and (f) the system with
$Q_{\mathrm{ion}} / Q_{\mathrm{rep}} = 0.125$. In order to see more clearly the
effects of the broken $C_4$ symmetry the averages are restricted to
configurations with $\phi \geq 0$ corresponding to the expected
response in a single domain sample. } 
\label{fig:CSF}
\end{figure}

We consider a $L\times L$ square cluster with even number of holes $N_h$, corresponding 
to $N_h/2$ TDs. Although we explored various fillings, for the sake of 
definiteness in the present Communication we report the results for the typical case 
$L=100$ and $N_h=300$, corresponding to a hole doping $n_h=0.03$. 

To characterize the broken symmetries, we define a nematic order parameter 
$\phi(T)$ [see Eq. (\ref{phiOP}) in Methods], which becomes different from zero 
when the $C_{4}$ rotational symmetry of the lattice is broken. We also define the polarization 
of the system as the normalized sum of all the TD moments projected on the $(1,1)$ and $(1,-1)$ 
preferred directions [cf. Methods, Eqs. (\ref{eq:p11}, \ref{eq:p1m1})]. A nonzero polarization 
in the system implies a breaking of inversion symmetry of the magnetic texture. Vector chiral
spin order is characterized by the chirality $\chi_{1,\pm 1}$ [see Eq. (\ref{chirality}) in Methods].
Finally, the charge and spin structure factors allow us to further characterize the various 
phases.

The Monte Carlo computations find at high temperature a classical
liquid of dimers, which tend to  form longer polymers as temperature
is lowered (Supplementary Figs.~6), and to align along the diagonal
directions, which are energetically favorable. Fig.~\ref{fig:snap}a
reports a snapshot of this high-temperature phase,  taken during the Monte Carlo evolution.  

\vspace{1 truecm}
\noindent
{\bf  Clean System.} For the clean system ($Q_{\mathrm{ion}}=0$) we
find that, when $T$ is low enough, the segments orient to  
form a state with $C_4$ symmetry breaking. As is clearly visible in
Fig.~\ref{fig:snap}b (see also Supplementary Fig.~5), associating 
the segments with ``polymers'', the low-$T$ phase corresponds to the so-called smectic order of soft matter~\cite{Chaikin_book}, 
in which the system has long-range positional order in one direction, with periodicity 
$\ell_c=1/q_c$ [$q_c$ being the magnitude of a characteristic
wave-vector in reciprocal lattice units (rlu)], but remains ``liquid'' in the 
other direction. This manifests as sharp (resolution limited) peaks in
the structure factor along the diagonal of the Brillouin zone  
(Fig.~\ref{fig:CSF}a), {\it i.e.}, perpendicular to the preferred
polymer direction, signaling long-range positional order. 
As  is shown in panel (e),  
the main spin peak is twenty times higher than the charge peak,
because the spectral weight of the latter is spread  
over a wider range of wave-vectors, due to the highly anharmonic
charge distribution. (Notice that both structure factors, as defined in
Methods, have the same normalization.)
Regarding CO,  this state  has the same symmetry as a diagonal stripe
state. However,  the charge is uniform along the stripe direction only  
after thermal fluctuations have been taken into account. In addition,
this state breaks inversion symmetry, {\it i.e.},  
TDs tend to point in the same direction, thus we call it ferrosmectic.

The ferro ordering associated with this and other phases is not trivial. Indeed, in contrast to dipoles  
on a cubic lattice in three dimensions, it would not occur if, for example, the TDs were arranged 
at fixed positions on a square lattice. This stems from the 2D dipole-dipole interaction, which is 
ferroelectric in a nearly head-to-tail configuration of the dipoles, but is antiferroelectric 
for a side-by-side configuration. In our model, the ferro tendency wins because the real Coulomb
interaction between the electrically charged holes favors short-range triangular arrangements of the segments, 
{\it i.e.}, segments in one row tend to face gaps in the neighbouring
rows as is clear in Fig. \ref{fig:snap}b (highlighted by white 
lines) and c, so that the side-by-side arrangements are rare. 
The colours in Fig.~\ref{fig:snap} show the phase of the local
staggered magnetization. In panels b-d the phase  
increases monotonically along one diagonal, indicating that these
phases have long-range vector chiral order, {\it i.e.},  
$\chi_{1,-1}\ne 0$ or $\chi_{1, 1}\ne 0$.

Upon further lowering the temperature, the ferrosmectic phase keeps the ferro ordering (and the vector spin 
chirality), but forms a Wigner crystal for $T \lesssim 10$\,K as shown
in Fig.~\ref{fig:snap}c. This
``ferrocrystal'' manifests as additional resolution limited 
off-diagonal peaks in the charge structure factor (as the one indicated with an
arrow in Fig.~\ref{fig:CSF}c) and which again signal long-range charge and spin
order.

{\bf Effect of Disorder.}
The properties of the phases change dramatically upon the introduction
of quenched disorder. The ferrosmectic and 
ferrocrystal peaks broaden and weaken very rapidly
(Fig.~\ref{fig:CSF}b,d,f), thus long-range positional 
order is lost and the ferrosmectic-ferrocrystal transition 
is smeared. Remarkably  long-range nematic and vector chiral order (accompanied by inversion symmetry breaking)  remain at finite disorder, and the phase becomes
the ferronematic state proposed in  Refs.~\cite{Seibold_PRB2013,Seibold_SR2014}.
Fig.~\ref{fig:CSF}f shows how CO is almost entirely destroyed by small disorder. Long range
spin order also is destroyed but short-range spin order, signaled by
incommensurate peaks, whose width is well resolved 
in our system size, persists. For $Q_{\mathrm{ion}}/ Q_{\mathrm{rep}}>
0.25$ even the broad incommensurate magnetic peaks disappear
(Supplementary Note 4 and Supplementary Figs. 6-8). This would contradict
experiments, we thus estimate $Q_{\mathrm{ion}}/Q_{\mathrm{rep}}<0.25$
in real systems. 
\begin{figure}[tb]
\includegraphics[width=9cm]{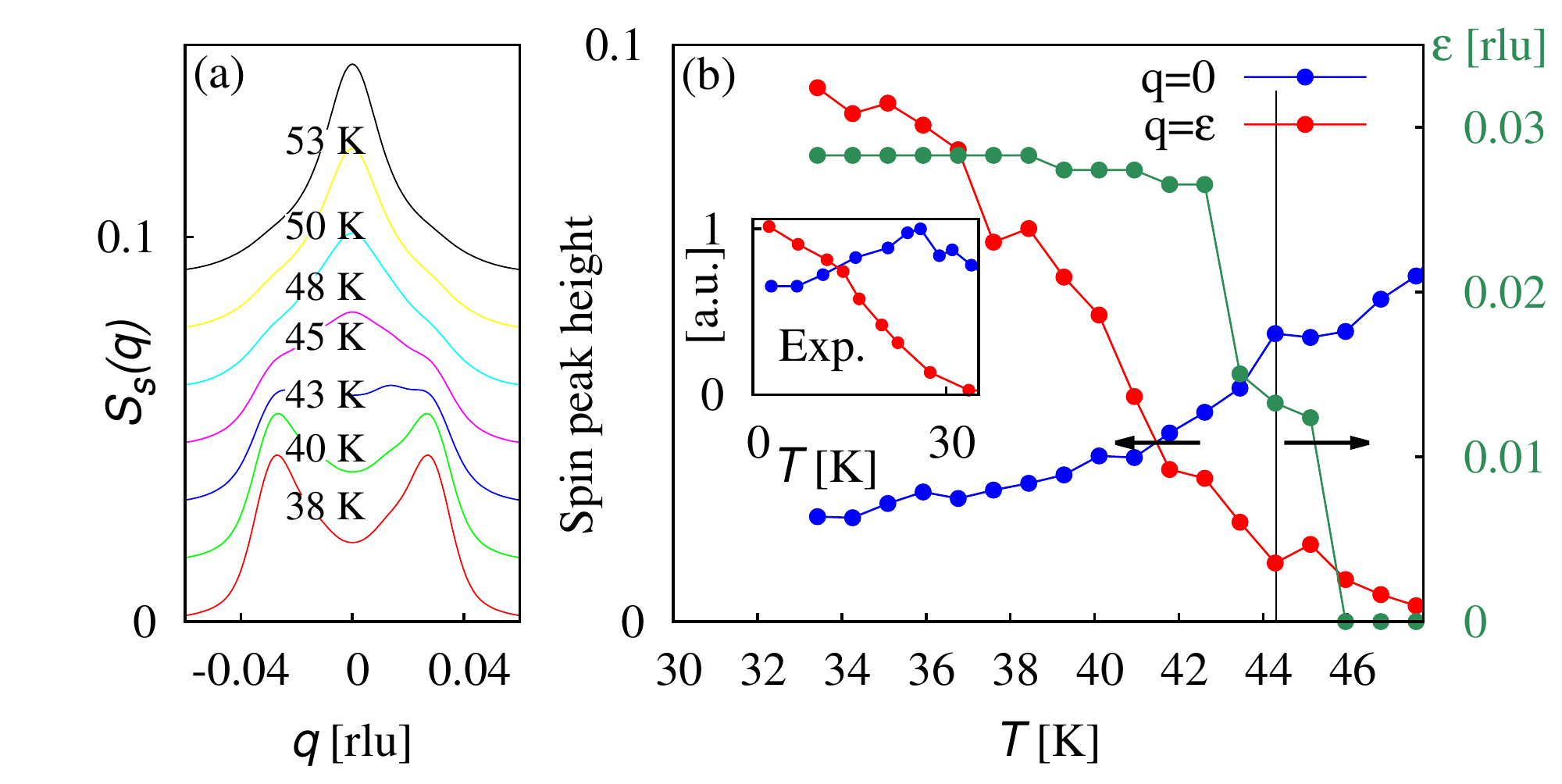}
\caption{{\bf Commensurate-incommensurate transition}
(a)~Diagonal cuts of the spin structure factor for different temperatures as a
function of momentum with $q$ defined as in Fig.~\ref{fig:CSF} and disorder $Q_{\mathrm{ion}} / Q_{\mathrm{rep}} = 0.125$. (f).  
The peaks have been convoluted 
 with a Gaussian (standard deviation $0.041$~[rlu]) to take into account a
 finite experimental resolution. (b)
Height of the structure factor shown in (a) at the commensurate
antiferromagnetic wave-vector (blue) and at the incommensurate
position with respect to the background (red) as a function of temperature. 
The green data (right scale)  shows the incommensurability as a
function of temperature. The vertical line marks the ferronematic transition. 
The inset shows the
experimental peaks height from Ref.~\cite{Drachuck_NatureComm2014} for doping
$n_h=0.0192$, slightly below the complete disappearance of static
antiferromagnetic order as reveled by muons. 
The evolution of the incommensurate peaks has
been shown to be continuous~\cite{Matsuda_PRB2002} across the critical doping $n_h=0.02$.  
}
\label{fig:OP_0.03}
\end{figure}     

The red solid circles in Fig.~\ref{fig:CSF}a-d show the 
vectors $\pm (\epsilon / \sqrt{2},-\epsilon / \sqrt{2})$\,r.l.u.
where $\epsilon$ is the magnetic incommensurability, {\it i.e.}, for the
orientation of Fig.~\ref{fig:snap}b,c magnetic peaks appear at 
 $q_{AF} \pm (\epsilon / \sqrt{2},-\epsilon / \sqrt{2})$\,r.l.u.,
with $q_{AF}=(0.5,0.5)$\,r.l.u. 
From all panels we see that the main magnetic peaks appear at half the
incommensurate wave-vector of the main charge peaks. At first 
sight, this relation, well known for spin collinear stripes~\cite{Lorenzana_PRL2002,sei04a}, is surprising here, since 
the incommensurability should be linked to the topological 
polarization~\cite{Seibold_PRB2013}. However, close inspection of 
Fig.~\ref{fig:snap}b reveals that each segment acts as an antiphase domain wall for the antiferromagnetic 
background, yielding jumps of the phase of the magnetic order parameter close to $\pi$ upon crossing the line of polymers. 
On the other hand, the phase is approximately constant in between two polymer rows. Thus, the magnetization behaves similarly to the case of a collinear stripe array. Spin canting produces small corrections to the 'factor of two' relation, which are below our momentum resolution to be visible in Fig.~\ref{fig:CSF}.

Raising the temperature at small disorder, the broadened spin  and
charge peaks gradually decrease (Fig.~\ref{fig:OP_0.03}a)  without 
any sign of a sharp transition
in the intensity (Fig.~\ref{fig:OP_0.03} and Supplementary Fig.~6,
respectively) as also observed experimentally at similar dopings (see inset of
Fig.~\ref{fig:OP_0.03}b and Ref.~\cite{Matsuda_PRB2002}).   
In contrast, studying the
polarization and nematic order parameter distribution we find that the
transition from the ferronematic to the melted polymers is of first 
order and remains sharp for our system size (see Supplementary Note 5).  Thus a
thermodynamic transition persists even in the presence of disorder.  
The thermodynamic transition temperature is signaled by a change of
behavior in the magnetic structure factor from commensurate to
incommensurate, providing a simple experimental tool to detect the
transition line (Fig.~\ref{fig:OP_0.03}a). This is because the
incommensurability $\epsilon$ is related to degree of polarization in
the system and thus acts as an order parameter~\cite{Seibold_PRB2013}.

\begin{figure}[tb]
\includegraphics[width=9cm]{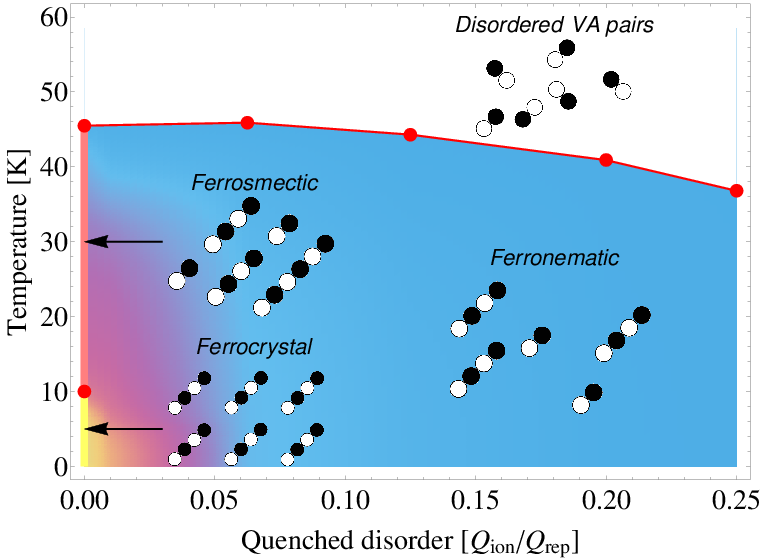}
\caption{Phase diagram as a function of temperature and disorder strength. The yellow (pink) thick line at 
zero disorder corresponds to ferrocrystal (ferrosmectic) long-range order. The yellow region is 
short-range ferrocrystal order while the magenta region corresponds to the short-range ferrosmectic 
order. At finite disorder, below the red line, the system has long-range ferronematic order 
(light blue region) while a polymeric liquid is found above the red line, up to the highest temperatures 
reached in our study.}
\label{fig:PD_dis}
\end{figure}

{\bf Phase Diagram.}
Fig.~\ref{fig:PD_dis} reports the phase diagram obtained from the above analysis. 
The ferrocrystal (thick yellow line) and 
ferrosmectic (thick pink line) phases are well defined only in the absence of disorder. At finite 
disorder, they survive as short-range ordered states. This is indicated by the yellow shade for the 
ferrocrystal and by the magenta shade for the ferrosmectic state. The light blue region is the 
long-range ordered ferronematic state, while the red line indicates the first-order transition to a liquid of
short polymers. We never find a purely nematic phase, characterized by a nonzero nematic order 
parameter but zero polarization and zero global vector chiral spin
order. The last phase, which is
allowed by our model, could possibly be stabilized in a different parameter regime, as an intermediate phase between 
the ferronematic and the disordered phase. 

Our results are in good 
qualitative agreement with the phase diagram obtained by completely
different methods in Ref.~\cite{Nie_PNAS2014}.  On the other hand, we find
an additional inversion symmetry breaking and we provide realistic
estimates of the parameters of the model, of the experimentally  
measurable structure factors and of the characteristic temperatures of
the transitions. 

The order of magnitude of the transition temperature
to a polarized state can be estimated using a mean-field
approximation. For dipoles in two-dimensions at random 
positions but with a non zero average dipole moment  $\langle p
\rangle$, the dipolar field can be computed using elementary
electrostatics to be $E_d=2\pi^2  \rho_s\langle p\rangle n_h/N_c 
$ where $\rho_s$ is the magnetic stiffness of the system and $N_c$ the
number of charges per segment. 
Assuming a mean-field approximation where the dipoles, of strength
$p_0\equiv\sqrt2 (N_c-1) k$, can fluctuate in
4 possible orientations ($\uparrow,\rightarrow,
\downarrow,\leftarrow $) we obtain the ordering temperature, 
\begin{equation}
  \label{eq:tc}
k_B T_c= (\pi p_0)^2 \rho_s \frac{n_h}{N_c} .
\end{equation}
With the present parameters and assuming $N_c\approx 4$ (Supplementary
Fig.~6), we find  $T_c\approx$ 196 K. This is of the correct order of
magnitude  ({\it i.e} much smaller than the original electronic scales)
taking into account that we have neglected the positional entropy which will
 reduce  $T_c$.

Both ferrosmectic and ferrocrystal charge orderings are not commensurate. Thus, they break a continuous [U(1)] 
symmetry in two dimensions. Even in the presence of infinitesimal
disorder strict long-range order is forbidden 
~\cite{Imry_PRL1975} and one finds quasi long- or short-range order. In contrast, the nematic order parameter breaks a discrete 
(Z$_2$) symmetry and is much more robust against
disorder. In our computations we have in addition vector chiral 
spin order (or equivalently a topological polarization) which also breaks a discrete (Z$_2$) symmetry, but which does 
not couple linearly to the local disorder, in contrast to the nematic
order parameter~\cite{Carlson_PRL2006,Phillabaum_NatComm2012}. General arguments indicate 
that the discrete symmetry breaking should be much more robust than
the breaking of  a continuous symmetry~\cite{Chandra_PRL1990,Nie_PNAS2014}, as we indeed
find. We expect the nematic order to behave similarly to the random field Ising model: lacking long-range 
order in a strictly 2D system, but ordered
within a correlation length which can be  
exponentially large for small disorder~\cite{Binder_ZP1983}, favoring a
crossover to three-dimensional long-range order in  the presence of a
small inter-layer coupling~\cite{Zachar_PRL2003,Phillabaum_NatComm2012}.

Since the ferronematic state has short-range spin order and long-range
vector chiral order (at ${\bf q}=0$ wave-vector), it can be identified with the chiral spin liquid believed
to take place in frustrated magnets~\cite{Kawamura_2002,Hikihara_PRB2008,Grohol_NatureMat2005}.

\vspace{1 truecm}
\noindent
{\bf\large  Discussion and conclusions} 

Our results allow us to rationalize several experimental findings, and imply some predictions which have 
not yet been tested. 

Experiments show that hole doping destroys commensurate
antiferromagnetic order much more rapidly than what would be expected
by site dilution~\cite{Cho_PRB1992,Coneri_PRB2010}.  Fig.~\ref{fig:CSF}f shows that 
this is explained by a small density of TDs.  
The ability of VA pairs to rapidly depress commensurate ordering was
noticed before~\cite{Aharony_PRL1988,Timm_PRL2000}, although these authors did not
consider the collective ordering of the dipoles.

Incommensurate spin scattering has been detected in the early days of
high-$T_c$~\cite{Cheong_PRL1991} and interpreted in terms of 
stripes~\cite{Tranquada_Nature1995,Tranquada_PRL1997}. 
However, stripes are associated with 
charge modulations which are extremely hard to measure, in contrast to
spin modulations. CO generally emerges associated with a structural
distortion close to $n_h= 1/8$ which can be controlled by codoping  with
Nd~\cite{Tranquada_Nature1995,Tranquada_PRL1997} 
or doping/codoping with
Ba~\cite{Fujita_PRL2002,Abbamonte_NatPhys2005,Hucker_PRB2011}. 
All these observation of CO are at doping
close to $n_h= 1/8$. The intensity of CO decreases strongly with
underdoping and extrapolates to zero
around $n_h\approx 0.09$~\cite{Hucker_PRB2011}. To the best of our knowledge,
incommensurate static charge order has
never been reported in the present, heavily underdoped, regime, in
contrast to incommensurate spin order~\cite{Matsuda_PRB2000}, which persists. 
This dichotomy is 
explained by our simulations which, while reproducing the
incommensurate spin ordering, show very weak 
charge-ordering peaks, barely emerging from the background noise, even for
weak disorder (Fig.~\ref{fig:CSF}f).  

Close to $n_h= 1/8$ magnetic Bragg peaks appear quite sharp and
often resolution 
limited~\cite{Tranquada_PRL1997,Kimura_PRB1999,Hucker_PRB2011} 
indicating long-range
order.  As doping is reduced static peaks are still observed but  
become broad with a well resolved width of the order of the incommensurability
indicating a correlation length of the order of the spin
periodicity~\cite{Matsuda_PRB2000,Hucker_PRB2011}.  This is in excellent agreement
with our magnetic structure factor in Fig.~\ref{fig:CSF}f. 
We interpret this feature as an indirect signature of long-range 
vector chiral spin order without long-range magnetic order {\it i.e.}
the ferronematic state we propose. 
For $n_h=0.03$, experimental magnetic peaks 
have been detected with incommensurability 
$\epsilon \approx 0.032$~\cite{Matsuda_PRB2002} in good agreement with our
computations yielding $\epsilon \approx 0.028$ at low temperature.

Neutron scattering experiments in Y-based materials have
shown~\cite{Hinkov_Science2008,Haug_NJOP2010} 
that the magnetic incommensurability as a function
of temperature behaves as an order parameter. Such a behavior is
naturally explained by our model, where the incommensurability, in  
the presence of weak or no CO, is closely linked to the topological
polarization~\cite{Seibold_PRB2013}, which is an order parameter (see
Fig.~\ref{fig:OP_0.03}).  Furthermore we propose that the temperature
at which the static magnetic 
structure factor changes from a double peak structure to a single peak
structure is a proxy of the thermodynamic critical
temperature below which long-range chiral spin order is established.

The transition from incommensurate behavior to commensurate behavior
has been observed also in the specific La-family we focus on in the present computations. 
Indeed experimental  
low-energy {\em inelastic} neutron scattering peaks as a function of
temperature reported in  Fig.~5a of Ref.~\cite{Matsuda_PRL2008} (see
also Ref.~\cite{Matsuda_PRB2000})
show the same behavior as we find for the static structure factor.  However the transition from 
two incommensurate peaks to an antiferromagnetic commensurate peak 
takes place around $55-100$~K. On the other hand {\em  quasistatic}
scattering shows a transition at around $20 -
30$~K (see Refs.~\cite{Wakimoto_PRB2000b,Matsuda_PRB2002,Drachuck_NatureComm2014} 
and inset of Fig.~\ref{fig:OP_0.03}b). Our
computations provide an energy integrated structure factor which is
expected to show the transition between the {\em inelastic} and {\em
  quasistatic} cases. Indeed we find  the  commensurate-incommensurate
transition at around 45~K fully consistent with the neutron scattering
measurements.   
 Such an agreement on the temperature scales and qualitative behavior 
further supports our identification of the low temperature state
observed in cuprates as a long-range-ordered ferronematic.  

Notice that, in 
contrast with the small ordering scales we find, the starting point
electronic Hamiltonian  
has bare electronic scales of the order of $3000$\,K or more.
This strong reduction of energy scales indicates that our multiscale
modeling has identified the correct dynamical variables of the
problem. Eq.~\eqref{eq:tc} shows that the energy scale is set by the
magnetic stiffness and the density. 
Notice also that the proposed thermodynamic transition occurs
at a temperature much lower than the 
pseudogap temperature ($\approx 300$\,K) which instead nearly extrapolates to
the N\'eel  temperature of the undoped sample~\cite{Ando_PRL2004}.

At even lower temperatures of the proposed ferronematic transition a so
called cluster spin glass state is observed consisting of strongly coupled 
clusters of spins with weaker coupling among 
clusters~\cite{Cho_PRB1992,Chou_PRL1993,Borsa_PRB1995,Niedermayer_PRL1998,Coneri_PRB2010,Wakimoto_PRB2000b}. 
The ferronematic state of
Fig.~\ref{fig:snap}d corresponds precisely to this physical
picture. 

Eq.~\eqref{eq:tc} predicts a linear relation between doping
and the ordering temperature. However for  $n_h>0.02$ one should 
take into account that the finite magnetic correlation
length, the expected weakening of the stiffness by doping and the
breakdown of the linear regime for the topological charges will lead
to a slowing down of the doping dependence. Interestingly for hole
content $n_h<0.02$ the temperature at which the cluster spin glass
state is observed is linear with 
doping~\cite{Chou_PRL1993,Borsa_PRB1995,Niedermayer_PRL1998} consistent with
our proposal. Numerically we find that the temperature of the
ferronematic transition increases approximately linear with doping as
$T_c\sim$~1500~K~$n_h$ compared with the experimental behavior 
$T_c\sim$~815~K~$n_h$.  Our larger sloop may be due to an overestimation
of $\rho_s$, a smaller $N_c$  and/or  dynamical effects which may
give an apparent shift of the transition.

In the presence of spin-orbit coupling, long-range vector chiral spin
order gives rise to a real electric polarization,  {\it i.e.},  
the system becomes an improper ferroelectric
~\cite{Cheong_NatureMat2007}. Unfortunately, this effect is hard to observe because,  
as soon as the system becomes metallic, it cannot support a finite
electric polarization. Notwithstanding, a 
finite ferroelectric polarization has been reported at low temperatures in 
oxygen~\cite{Viskadourakis_PRB2012} and Li~\cite{Viskadourakis_2014} 
doped La$_2$CuO$_4$, the samples having a strongly insulating character. The fact that the effect appears independently 
of the dopant, and that the remnant polarization can be oriented along different axes with external fields, clearly 
points to a magnetic origin of the ferroelectric polarization. Furthermore, more recent experiments show a clear 
correlation between magnetoelectric effects and stripe orientation in Sr doped La$_2$NiO$_4$, suggesting that 
stripe effects are involved~\cite{Panagopoulos_PrivComm}. Experiments at finite frequencies suggest that inversion 
symmetry breaking sets in at temperatures higher than the temperatures at which the sample is insulating enough to 
support a static polarization. All these experiments support our
conclusion that underdoped cuprates show long-range vector chiral spin
order. 

A possible test to our model would require second harmonic
generation to detect inversion symmetry breaking in non-insulating
samples. We predict that in the ferronematic phase the inversion
symmetry breaking should track the behavior of the incommensurability
as a function of temperature. This relation, however,  
will break down in the collinear stripe phase found around $n_h=1/8$. 

With the present method we cannot access quantitatively the crossover
to collinear stripes.  In this regime, the mapping to the
Coulomb gas breaks down due to nonlinear  
effects. However, one can anticipate that the average length of the segments 
will keep growing with doping, leading to a concomitant increase of the 
ferrosmectic correlation length. According to our findings, the disorder induced by the dopants will partially 
counteract this increase, but the associated impurity potential will
also be progressively screened, opening the possibility that segments
coalesce into stripes with long-range order and narrow magnetic
peaks. 
  
We thus propose that underdoped cuprates have a long-range broken
symmetry state at low doping. This puts the cuprate phase diagram into
the same class of phase diagrams of a wide class of materials~\cite{Taillefer_ARCMP2010} 
in which unconventional superconductivity emerges from a phase characterized by
real-space electronic long-range order. 

\vspace{1 truecm}
\noindent
{\bf\large Methods}

{\bf  Model.}
Treating the single-band Hubbard model within a Gutzwiller approximation, a single hole in the 
antiferromagnetic background is found to form a spin polaron, while two holes tend to occupy the cores of a spin 
V and A that attract each other, thereby lowering their energy. The
long-range part of this texture is treated  
using generalized elasticity~\cite{Chaikin_book} and exploiting the
correspondence between a spin vortex and a 2D Coulomb  
charge~\cite{Minnhagen_RMP1987}. In the absence of disorder and holes the
magnetic correlation length is expected to be very large but finite
due to thermal fluctuations. This provides a natural cutoff at a
distance $\lambda$ for the 
long-range interactions between topological charges at large
distances~\cite{Aharony_PRL1988,Timm_PRL2000}. 
Therefore, the interaction energy is well described by
\begin{eqnarray}\label{eqn:topological_int}
V_{lr}(r) &=& \rho_s k_1 k_2 \int_0^{2\pi}{\mathrm d}\theta
\int_0^{\infty}{\mathrm d}q  \,\frac{q\,{\mathrm e}^{iqr\cos\theta}}{q^2 + \lambda^{-2}}
\nonumber\\
&=& 2 \pi \rho_s k_1 k_2 K_0\left(\frac{r}{\lambda}\right),
\end{eqnarray}
where  
$k_{1,2}=\pm k$ are the topological charges (in our case, $k=0.8$) and $K_0$ is the zeroth order 
modified Bessel function, which reproduces the logarithmic interaction at short/intermediate 
distances ($r \lesssim\lambda$) and decays exponentially at long
distances ($r \gg \lambda$).  We have checked
that changing $\lambda$ the results are substantially the same, as
long as it remains larger than the typical distance between the  
segments. In the simulations we take
$\lambda$ of the order of the system size for numerical convergence  
purposes. 

While Eq.~\eqref{eqn:topological_int} reproduces well the energy of Gutzwiller calculations for 
the single band Hubbard model at large distances, as expected, it fails at short distances where 
the short-range physics of the Hubbard model becomes
relevant. Therefore, we also include short-range terms extrapolated
from the Gutzwiller calculations (Supplementary Note 1). 

Furthermore, each topological charge arising from the spin texture corresponds to a positive 
electrically charged hole in the CuO$_2$ planes of the doped cuprate. Therefore, our model includes
also the three-dimensional Coulomb repulsion potential which, for two holes at a
distance $r$, is parametrized as
\ie
v_{hh}(r) = \frac{Q_{\mathrm{rep}}}{r}. 
\label{hh_repuls}
\fe 
Here $Q_{\mathrm{rep}}$, incorporating, e.g., the static dielectric constant, represents the strength of the 
repulsion and is another parameter of our model.
We fix $Q_{\mathrm{rep}} / a = 49$\,meV, where $a$ is the
in-plane lattice constant, so that the average number 
of holes in a polymer, for very low density is $N_c \approx 2$.
As the density increases, this number tends to increase 
too~\cite{Seibold_PRB2013}, yielding the results of Supplementary Fig.~6
for the present density ($n_h=0.03$).

Finally, the charged holes doped into the CuO$_2$ planes leave back negative countercharges. 
For instance, in \LSCO, which we take as a prototype cuprate, negative Sr ions randomly replace La atoms 
between two consecutive planes. We therefore introduce disorder, generating a random distribution of 
point-like negative charges, one for each positive hole in the plane, 
which act as pinning centers for the carriers in the plane
(Supplementary Note 2). 
The ions are located out of plane, at a distance $\bar d\approx 0.58
a$, from the center of the in-plane plaquette. Each impurity interacts with the holes in the plane through an attractive 
three-dimensional Coulomb potential
\ie
v_{\mathrm{ion}} = - \frac{Q_{\mathrm{ion}}}{d} ,
\fe 
where $d$ is the distance between the hole  and the
impurity, and the strength of the interaction $Q_{\mathrm{ion}}$
measures the intensity of disorder. This procedure thus produces a
disordered potential in the plane in which holes and their associated
topological charges move. We show one
realization of the impurity potential in Supplementary Fig.~4.

{\bf  Characterization of the phases.}
To characterize $C_4$ rotation symmetry breaking, we introduce the nematic order parameter
\begin{eqnarray}
\phi& =& \frac{1}{N_h} \sum_{\br_i} \langle n(\br_i) n(\br_i + \hat{x} + \hat{y}) \nonumber\\
&-& n(\br_i + \hat{y}) n(\br_i + \hat{x}) \rangle, \label{phiOP}
\end{eqnarray}
where $n(\br_i)$ is the number of holes on the site labelled by $\br_i$, and $\hat{x},\hat{y}$ are
unit vectors along the corresponding directions of our 2D square lattice. The angular
brackets imply thermal average, $ N_h$  denotes the total number of charges, and
the sum runs over all the lattice sites. To characterize
inversion symmetry breaking, we introduce the polarization $\bP = (P_x,P_y)$, as the normalized sum of all 
the TDs. Since in our model diagonal polarizations are favoured, we introduce the components 
\begin{eqnarray}
P_{(1,1)} &=& (P_x+P_y)/\sqrt2\, , \label{eq:p11}\\ 
P_{(1,-1)} &=& (P_x-P_y)/\sqrt2 \label{eq:p1m1}\,.
\end{eqnarray}
Vector chiral spin order is characterized by the parameter
\begin{equation}
\chi_{1,\pm 1}\equiv\frac{1}{\sqrt{L^2}}\sum_{\br_i}
\langle [ \bs(\br_i) \times
\bs(\br_i+\hat{x}\pm\hat{y})] \cdot \hat{z} \rangle,
\label{chirality}
\end{equation}
where $\bs(\br_i)$ is the local spin density.

Our Monte Carlo calculations also yield the thermal averages of the static charge (c) and spin 
(s) structure factors, $S_c(\bq) = (1-\delta_{\bq,0}) {|K_c(\bq)|^2}/{L^2}$ 
and $S_s(\bq)={|\bK_s(\bq)|^2}/{L^2}$, where
\begin{eqnarray}
K_c(\bq) &=& \frac{1}{\sqrt{ N_{h}}}\sum_{\br_i} \exp(i \bq \cdot
\br_i) \;  n(\br_i) \\
\bK_s(\bq) &=& \frac{1}{\sqrt{L^2}}\sum_{\br_i} \exp(i \bq \cdot \br_i) \; \bs(\br_i).
\end{eqnarray}
With these definitions and using unit-length spins, the structure
factors have the same normalization and satisfy $\sum_\bq  S_{s,c}(\bq)=1-n_h$. 

The ferronematic-ferrosmectic crossover in Fig.~\ref{fig:PD_dis} was characterized
analyzing the height of the main charge peak as a function of
temperature.

{\bf Monte Carlo analysis.}
We carried out Monte Carlo calculations exploiting the parallel tempering technique. To 
analyze the spin degrees of freedom for a given configuration, we attach to each topological charge the 
structure of a (anti)vortex in the spin background and we perform a linear superposition, allowing 
then each spin to relax according to the XY Hamiltonian~\cite{Minnhagen_RMP1987}. 
Supplementary Fig. 1b reports an example for the case of two VA pairs aggregated in a 
four site segment. Further examples with a detailed view of the segments, of the corresponding
relaxed spin structures and of the resulting spin currents are
reported in Supplementary Note 3 Supplementary Fig.~5. 

The temperature step of our simulations is $0.4$\,K ($0.8$\,K) for the 
clean (disordered) case. 
To better determine the various phases, at each temperature we construct a
histogram over the Monte Carlo history 
defined on a three-dimensional grid spanned by the 
order parameter $(\phi,P_{(1,1)},P_{(1,-1)})$ (see Ref.~\cite{Capati_PRB2011}). 
The probability for a value $(\phi,P_{(1,1)},P_{(1,-1)})$ 
of the order parameter is given by the Boltzmann factor
$\sim\exp[-F(\phi,P_{(1,1)},P_{(1,-1)}) /(k_B T)]$, where $F$ is the free 
energy. Finding the position  of the maximum of the histogram [which is a point in three-dimensional 
space $(\phi,P_{(1,1)},P_{(1,-1)})$] is then equivalent to minimize the free energy and identifies the
stablest phase. This yields sharper transitions than following the thermal average which, with our 
accessible system sizes, often is not large enough to resolve closely
separated transitions.  More details are given in Supplementary
Figs.~9 and 10 and Supplementary Note 5. 

{\bf Acknowledgments.} We thank M. H\"ucker, L. Benfatto,
C. Castellani, A. Pelissetto,  C. Pierleoni, R. de Renzi, P. Carretta, 
S. Sanna, A. Keren, G. Drachuck, J. Hellsvik and S. Kivelson  for valuable
discussions and  insightful comments on the manuscript.
S.C., C.D.C. and M.G. acknowledge financial support from the Progetti
AWARDS of the Sapienza  
University, Project n. C26H13KZS9. J.L. acknowledges support from the
Italian Institute of 
Technology through seed project NEWDFESCM, from the Simons Foundation
and the hospitality of  
the Aspen Center for Physics. G.S. acknowledges financial support from
the Deutsche  
Forschungsgemeinschaft.

{\bf Author Contributions.} S.C., C.D.C, M.G., G.S and J.L conceived the research. J.L. supervised the research. G.S and M.C realized the Gutzwiller approximation computations. M.C. derived the effective model, realized the Monte Carlo computations and analyzed the data. M.C., M.G and J.L. drafted the manuscript. All the authors contributed to the interpretation of the results and the writing. 

{\bf Competing financial interests.}
The authors declare no competing financial interests.


\end{document}